\documentclass[doublecol,linenumbers]{epl2}

\newcommand{\vc}[1]{\mbox{\boldmath $#1$}}
\usepackage{amsmath}
\usepackage{amssymb}
\usepackage{graphicx}
\usepackage{dcolumn}
\usepackage{bm}
\usepackage{here}
\usepackage{color}
\usepackage{ulem}
\usepackage{MnSymbol}

\newcommand{\red}[1]{\textcolor{red}{#1}}
\newcommand{\blue}[1]{\textcolor{blue}{#1}}
\definecolor{mygreen}{rgb}{0.2,0.8,0.2}
\newcommand{\green}[1]{\textcolor{mygreen}{#1}}

\def\nn{\nonumber}
\def\bH{\begin{Huge}}
\def\eH{\end{Huge}}
\def\bL{\begin{Large}}
\def\eL{\end{Large}}
\def\bl{\begin{large}}
\def\el{\end{large}}
\def\beq{\begin{eqnarray}}
\def\eeq{\end{eqnarray}}

\def\eps{\epsilon}

\def\omg{\omega}

\def\e{{\rm e}}

\def\pr{\partial}

\def\>{\rangle}
\def\<{\langle}

\def\msd{\Delta J^2}

\def\H{\hat{H}}
\def\U{\hat{U}}

\def\v{\hat{v}}
\def\J{\hat{J}}
\def\p{\hat{p}}
\def\q{\hat{q}}
\def\corr{C{\rm r}}

\def\vq{\hat{\vc{q}}}
\def\vp{\hat{\vc{p}}}

\def\U{\hat{U}}
\def\dimN{N_{\rm dim}}

\def\Sm1{S_m^{(1)}}
\def\rhom1{\rho_m^{(1)}}
\def\tlife{\tau_L}
\def\avrtlife{\langle \! \langle \tau_L \rangle \! \rangle}
\def\lifetime{{\rm lifetime }}

\title{Measuring lifetime of correspondence with classical decay of correlation in quantum chaos}

\author{
Fumihiro Matsui{1}\thanks{E-mail: \email{fumihiro-matsui[at]xnea.net}} 
\and Hiroaki S. Yamada{2}\thanks{E-mail: \email{hyamada[at]uranus.dti.ne.jp}}
\and Kensuke S. Ikeda{3}\thanks{E-mail: \email{ahoo[at]ike-dyn.ritsumei.ac.jp}}
}
\shortauthor{F. Matsui, H. S. Yamada, K. S. Ikeda}

\institute{
\inst{1}{Demartment of Physics, College of Science and Engineering, Ritsumeikan University,Noji-higashi 1-1-1, Kusatsu 525-8577, Japan}\\
\inst{2}{Yamada Physics Research Laboratory, Aoyama 5-7-14-205, Niigata 950-2002, Japan}\\
\inst{3}{College of Science and Engineering, Ritsumeikan University, Noji-higashi 1-1-1, Kusatsu 525-8577, Japan}
}


\date{\today}
\abstract{
A very weakly coupled linear oscillator is proposed as a detector for 
observing time-irreversible characteristics of a quantum system, and 
it is used to measure the lifetime during which a classically chaotic 
quantum system shows decay of correlation.  Except for a particular 
case where the lifetime agrees with the conventional Heisenberg time, 
which is proportional to the Hilbert space dimension $N$, it is in 
general much longer: the lifetime increases in proportion to the product 
of $N$ and the number of superposed eigenstates, and is proportional 
to $N^2$ in the case of full superposition.
}

\begin{document}
\maketitle

A classical chaotic system can be the simplest origin of irreversibility \cite{prigogine}. 
Even a low-dimensional classical system exhibits mixing in a fully chaotic state, 
which means the loss of memory in the sense of decay of autocorrelation.  
Its quantum counterpart could be the minimal quantal unit exhibiting loss of memory 
and time-irreversibility.
Indeed, normal diffusion \cite{casati,fishman}, energy dissipation  \cite{ikedadissip},
energy spreading \cite{cohen} and many apparent features of 
irreversible phenomena can be realized in classically chaotic quantum systems 
with a small number of degrees of freedom. 

Appearance of the loss of memory in isolated quantum systems is also a severe 
problem limiting the performance of quantum computation algorithms \cite{quantumcomputation}.
If the chaotic region is bounded in the phase space and thus 
the effective dimension $\dimN$ of the subspace of Hilbert space relevant 
for the chaotic region is finite, the decay of correlation in quantum system
can be observed only on a finite time scale.
Even if the system is unbounded and $\dimN$ is infinite,
persistent coherence and localization properties inherent 
in a quantum system may prevent the system from complete decay of correlation 
\cite{casati,fishman}. 
However, the multidimensional unbounded quantum chaos systems can mimic classical
chaotic dynamics \cite{quantumclassical} and recover 
classical normal diffusion implying the complete decay of correlation 
(the Anderson transition) \cite{casatianderson,delande}.

Irreversibility in quantum system has been explored directly by 
numerical time-reversal experiments \cite{timerevexp,timerevexp-ballentine,benenti}. 
In particular, it has been extensively investigated by many investigators 
in the context of fidelity \cite{fidelity1,fidelity2}. 
The time scale on which quantum chaos can show exponential sensitivity is 
quite short and is up to the Ehrenfest time proportional to $\log(\dimN)$  
at the most \cite{bermanehrenfest}, but the time scale on which the decay 
of correlation is observed is much longer and is said to be as long as 
the Heisenberg time, which is proportional to $\dimN$ \cite{fidelity2}.

The main purposes of the present article are first to propose a general 
method for observing characteristics of time-irreversibility related to the 
mixing and the decay of correlation in isolated quantum systems, and next, 
to apply the method to classically chaotic quantum systems
by measuring the maximal time beyond which classical decay of correlation 
can no longer be observed. This time is the lifetime during which quantum 
systems can mimic classical time-irreversibility. 
We hereafter call it the lifetime of correspondence with classical decay 
of correlation, or simply {\it lifetime}.

Even in classical dynamics, the decay of correlation in chaotic systems
is realized only for ideally chaotic systems such as a C-system or a K-system. 
We thus  limit our consideration to quantum systems which show an ideal 
Markovian chaotic behavior in the classical limit. Even for 
such a limited situation, the \lifetime introduced above has not been 
investigated quantitively, in particular, for systems which are finitely 
bounded in the phase space and so can not exhibit explicit diffusive behavior. 

The fidelity method is a powerful tool extracting directly time-irreversible 
characteristics of quantum system \cite{fidelity2}. However, the fidelity method 
is not convenient for the purpose of observing irreversible characteristics with 
adequate numerical accuracy over an extremely long time scale without 
disturbing the dynamics of the examined system.

We introduce a linear oscillator as a detector for observing 
time-irreversible characteristics of a quantum system. 
If the object system is classically chaotic, 
the oscillator "converts" the motion of the object system 
to a Brownian motion in the homogeneously and infinitely 
extended linear oscillator's action space. 
Let the Hamiltonian of the examined object system ``S'' be $\H(\vp,\vq,t)$
~($\vq$ and $\vp$ are coordinate and momentum vector operator, respectively)
and let it be very weakly coupled with the linear oscillator ``L'', which is 
represented by angle-action canonical pair operators $\hat{\theta}$ and 
$\J=-i\hbar d/d\theta$ with the Hamiltonian $\omega \J$ of the frequency $\omega$. 
The total Hamiltonian reads 
\begin{eqnarray}
\label{hamil}
  \H_{tot}(t) = \H(\vp,\vq,t) + \eta \v(\vp,\vq)g(\hat{\theta}) + \omega \J, 
\end{eqnarray}
where $\v$ is a Hermitian operator, and $g(\theta)$ 
is a $2\pi$-periodic function with null average \cite{shepe83}, which is usually 
taken to be $g(\theta)=\cos(\theta)$. $\J$ has the eigenvalue $J=j\hbar~(j\in {\bf Z})$ 
for the eigenstate 
$\<\theta|J\>=\e^{-iJ\theta/\hbar}$
because of the $2\pi$-periodicity 
in $\theta$-space. 
Technically it is convenient to impose the action periodic boundary condition 
identifying $J=L\hbar$ with $J=-L\hbar$, which quantizes $\theta$ as 
$\theta_k=2\pi k/(2L)~~(k\in {\bf Z})$, where $-L<k\leq L$, 
and so $\<\theta_k|J\>=\e^{-iJ\theta_k/\hbar}/\sqrt{2L}$. 
From the Heisenberg equation of motion it immediately follows that
$\J(t)-\J=\eta \int^t_0 \sin(\omega s+\theta)\v(\vp(s),\vq(s))ds$. 
\red{
Here we mainly consider the case of $\omega\neq 0$ in order to remove the DC component 
of $\v(\vp(s),\vq(s))$ which exists in general and hinders irreversibility-related information, 
but the particular case $\omega=0$ is also important and it is also discussed later. 
}
If the fluctuating
part of $\v$ has an stationary auto-correlation with vanishing tail, 
$\J(t)$ exhibits a Brownian motion in the action space. An ideal example of 
a dynamical system realizing the above feature is a classical chaotic 
system such as a C-system or a K-system, which have the ideal Markovian property. 
For such systems a stationary normal diffusion is realized in the $J$ space 
as a typical irreversible behavior.

To be concrete, we hereafter confine ourselves to the case that the system 
$\H(\vp,\vq,t)$ is a kicked rotor (KR) whose classical version exhibits ideal chaos
with the Markovian property. However, a one-dimensional quantum kicked rotor 
can not mimic classical chaotic motion because of the quantum interference effect 
\cite{casatianderson} and can not be an ideal example. 
We therefore take as the object system $S$ the coupled kicked rotor (CKR) 
with the Hamiltonian $H(\vp,\vq,t)=\sum_{i=1}^2[\p_i^2/2+V(\q_i)\Delta(t)]+
\eps V_{12}(\q_1,\q_2)\Delta(t)$, as a sample system S, where 
$\Delta(t)=\sum_{\ell}\delta(t-\ell T)$
\blue{
represents the $\delta$-functional kick with the period $T$. 
} 
It can mimic well the classical chaotic motion \cite{quantumclassical}, 
which is induced by the entanglement between the two constituent systems \cite{couplerotorentangle}.
The CKR is observed at the integer multiple of the fundamental period $T$ as $t=\tau T$ where
$\tau \in {\bf Z}$. Then the one step evolution of the CKR 
from $t(=\tau T+0)$ to $t+T$ is described by the unitary operator 
\beq
\U=\e^{-i[V(\q_1)+V(\q_2)+\eps V_{12}(\q_1,\q_2)]/\hbar}\e^{-i[\p_1^2+\p_2^2]/2\hbar}. 
\eeq
\blue{
Let the Heisenberg representation of the operator $\hat{X}$ be $\hat{X}(\tau)=\U^{-\tau}\hat{X}\U^{\tau}$, 
then the time evolution by a single operation of $\U$ yields the mapping rule 
\beq
\vq(\tau+1)&=&\vq(\tau)+\vp(\tau)\\
\vp(\tau+1)&=&\vp(\tau)-\bigg[\frac{\pr}{\pr\vc{q}}(V(\vc{q})+V_{12}(\vc{q})})\bigg]_{\vc{q}=\vq(\tau+1).
\eeq
}
Each KR, say KR1 and KR2, is defined in the 
bounded phase space $(q_i,p_i)\in [0,2\pi]\times[0,2\pi]$ and 
$\hbar=2\pi/N_i$, where $N_1=N_2$ is a positive integer, 
and the dimension of the Hilbert space for the CKR is $N=N_1N_2=N_1^2$. 
This is the most important parameter which controls the lifetime, 
as will be shown later on. Since our system is bounded, 
it can not exhibit diffusive motion. 
But a diffusive motion may be realized in the additional degree of freedom of L.
As for the specific form of $V$
we take mainly the Arnold's cat map $V(\q_i)=-K\q_i^2/2$ and also the standard 
map $V(\q_i)=K\cos \q_i$ (defined in the bounded phase space $[0,2\pi]\times[0,2\pi]$)
with the interaction $V_{12}(\q_1,\q_2)=\cos(\q_1-\q_2)$. 
Hereafter, we couple L with S via only the first KR (KR1) as 
$\hat{v}(\bm{\p},\bm{\q})=\hat{v}(\p_1)=\sin(\p_1)$ 
so that the entanglement between the KRs may sensitively be reflected in the dynamics of L, 
and take the initial states of L and S as $|J=0\>$ and $|\Psi_0\>$, respectively. 

\blue{
Coupling L with S by the scheme of the Hamiltonian (\ref{hamil}), 
the Heisenberg equation of motion for the canonical pair operator 
($\hat{\theta},\J)$ yields the mapping rule for $\J(\tau)$
as 
}
\blue{
\begin{eqnarray}
\label{Jmap}
\J(\tau+1)-\J(\tau)=2\eta\frac{\sin(\omega T/2)}{\omega}\sin(\omega T\tau+\hat{\theta}+\omega T/2)\v_{\tau},
\end{eqnarray}
}
\blue{
where $\v_{\tau}=\U^{-\tau}\v(\vp,\vq)\U^\tau=\v(\vp(\tau),\vq(\tau))$. 
Recall that $\v_{\tau}$ varies only when the kick is applied
at $t=\tau T$, and in our case $\v_{\tau}=\sin(\p_1(\tau))$, 
and $\hat{\theta}(\tau) = \hat{\theta} + \omega T \tau$.
We reduce the coupling strength $\eta$ weak enough such that 
the backaction from L to S is negligible, then the motion of 
$\vp(\tau)$ and $\vq(\tau)$ is independent of L.
The bracket $\<...\>$ denotes 
the expectation value with respect to the initial condition $|0\>=|\Psi_0\> \otimes |J=0\>$.
Then the mean square displacement (MSD) of action of L, 
namely $\msd(\tau)=\<(\J(\tau)-\<\J(\tau)\>)^2\>$ is equal to $\<\J^2(\tau)\>$,
because $\<\J(\tau)\>=\<\J(0)\>=0$ (note that the RHS of Eq.(\ref{Jmap}) vanishes by the summation over $\theta$), 
and further $\<\J^2(\tau)\>= \<(\J(\tau)-\J(0))^2\>$.
Therefore the MSD can be expressed by using Eq.(\ref{Jmap}) recursively to leads to 
}
\begin{eqnarray}
\label{J2corr}
  \msd(\tau) = \sum_{s=0}^{\tau-1} D_\omega(s),~ D_\omega(\tau)=D\sum_{s=-\tau}^\tau \corr_\tau(s)\cos(\omega Ts),
\end{eqnarray}
\blue{ 
where $D =2\eta^2\sin^2(\omega T/2)/\omega^2$ and
\begin{eqnarray}
\label{autocorr}
  \corr_\tau(s)=(\<\Psi_0|\v_{\tau} \v_{\tau-s}|\Psi_0\> + c.c.)/2. 
\end{eqnarray}
}
The autocorrelation function satisfies $\corr_\tau(-s)=\corr_\tau(s)$. 
Scaled by $\eta^2$ the result does not depend on $\eta$ if it is taken 
sufficiently small.

Equation (\ref{J2corr}) works in both quantum and classical dynamics. 
The MSD is the sum of $D_\omega(\tau)$, which is nothing more than the finite time Fourier 
component of the autocorrelation function.
For classical KR having the ideally chaotic property
the autocorrelation function decays because of the Markovian nature of 
the classical chaotic dynamics, which is the indicator of chaotic irreversibility. 
Decay of autocorrelation function makes the finite time Fourier component 
\red{
$D_\omega(\tau)$ converge to the classical diffusion constant $D_\omega^{(cl)}$
as $\tau\to\infty$. 
}
\red {
The quantum motion can follow its classical counterpart
at least in the initial stage, if we prepare the ensemble of classical 
initial points 
such that its probability distribution agrees with the quantum probability 
distribution of the initial state $|\Psi_0\>$. 
Under such situation, the quantum autocorrelation function decays following
its classical counterpart temporally and $D_\omega(\tau)$ follows
the classical $D_\omega^{(cl)}$. Therefore, $\msd(\tau)=D_\omega^{(cl)}\tau$ in the initial stage.
However, $D_\omega(\tau)$ finally goes down to 0 on time average, because the 
quantum autocorrelation function is the 
sum of the finite number, i.e. $N$, of trigonometric function.
This means that $\msd$ saturates 
at a finite value $\msd_\infty$ as $\tau \to \infty$. 
} 
$D_\omega(\tau)$ deviates from the classical value $D_\omega^{(cl)}$ 
and eventually approaches toward 0. This means that
the decay of autocorrelation function mimicking the 
classical irreversibility disappears
at a characteristic finite time which we define as
the \lifetime $\tau_L$.
\begin{figure}[htbp]
\begin{center}   
\includegraphics[height=5cm]
{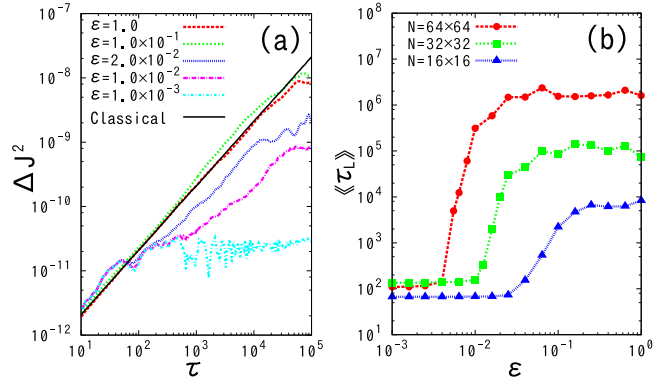}
\caption{\label{Fig1}(a) 
Time evolution of $\msd$ for various values of $\eps$, 
where $N=32\times32$. Here $K=10$, $\eta=10^{-6}$ and 
\red{$T=1$ throughout this paper.}
(b) Sharp transition observed by the averaged \lifetime as a function of $\eps$
for $N=16\times16,~32\times32$, and $64\times64$, where $\hbar=2\pi/N_1=2\pi/\sqrt{N}$.
Note that the averaged \lifetime above the critical $\eps$ is much larger 
than the order of magnitude of $N$.
}
\end{center}
\end{figure}

In practice we define the \lifetime $\tlife$ as the time beyond which
the stationarity of diffusion of L is lost: 
We introduce the temporal diffusion exponent $\alpha(\tau)$
at $\tau$ such that $\msd(s) \propto s^{\alpha(\tau)}$
for the appropriate interval of $s$ in the logarithmic scale.
We define $\tlife$ as the first time step 
at which $\alpha$ deviates from $1$ such that $|\alpha(\tlife)-1|>r$,
where the threshold value $r(<1)$ is chosen as follows: 
the diffusion exponent fluctuates in time around 1 from 0 to 2; $\alpha=0$ implies
the tendency toward the saturation, while $\alpha=2$ means tendency to exhibit
a ballistic motion due to the temporal quantum resonance. 
Therefore, it is reasonable to choose $r=0.5$.

The \lifetime thus defined in general varies very wildly with the choice of $|\Psi_0\>$.
In order to eliminate such accidental fluctuations, 
we add a classically negligible small term such as $\xi_{iR}\cos(\q_i-q_{iR})$ 
of $\xi_{iR}\sim O(\hbar)$ to the potential $V(\q_i)$, and take the average 
of $\tlife$ with respect to the ensemble of the phase parameter $q_{iR}~(i=1,2)$.
\footnote
{ 
\red{ 
Such additional potential term does not change classical chaotic dynamics any more, 
but the lifetime of quantum counterpart fluctuates wildly with the value of $q_{iR}~(i=1,2)$. 
Therefore its average is significant.
}
} 
We refer to it 
hereafter as the average \lifetime $\avrtlife$
\footnote{
It might seem adequate to take an average over 
an ensemble of $\omega$. But we do not take such an average here, because
the particular case $\omega=0$ is demonstrated. 
\red{ 
The averaging with respect 
to $\omega$ results in a similar effect as the averaging over $q_{iR}~(i=1,2)$.
} 
} 

Figure \ref{Fig1}(a) shows a typical example of $\msd$ vs $\tau$ at various
values of $\eps$ increased very slightly from $0$. 
These examples are obtained for 
the initial state $|\Psi_0\> = |p_1=N_1/2\> \otimes |p_2=N_2/2\>$, 
which is composed 
of 
\green{ 
all the eigenstates of the evolution operator $\U$
} 
 with almost equal weights and random phases.
The classical result is shown by a black line 
exhibiting an ideal diffusion with almost the same diffusion constant.
The quantum motions can follow the classical normal diffusion only 
within a finite time scale. They all deviate from the classical diffusion line and
tend to saturate at finite levels. Note that the time at which the 
deviation from classical normal diffusion occurs increases with $\eps$.
Indeed, as is shown in Fig.\ref{Fig1}(b), the averaged \lifetime increases drastically as 
$\eps$ exceeds the classically negligible characteristic value $\eps_c$ which is 
proportional to $\hbar^2$. 
Figure \ref{Fig1} implies that $\eps_c$ is the threshold above 
which quantum-classical correspondence is well realized and the \lifetime is maximally enhanced. 
This situation corresponds to the recovery of classical chaotic diffusion
in the coupled quantum standard map with unbounded momentum space \cite{quantumclassical}. 
We note that the \lifetime $\sim 10^6$ for $N=64\times 64$ greatly
exceeds the Heisenberg time $\propto N$.
We are interested in the nature of the \lifetime in such a situation 
that the quantum-classical correspondence is well achieved.

\begin{figure*}[htbp]    
\begin{center}
\includegraphics[height=5cm]
{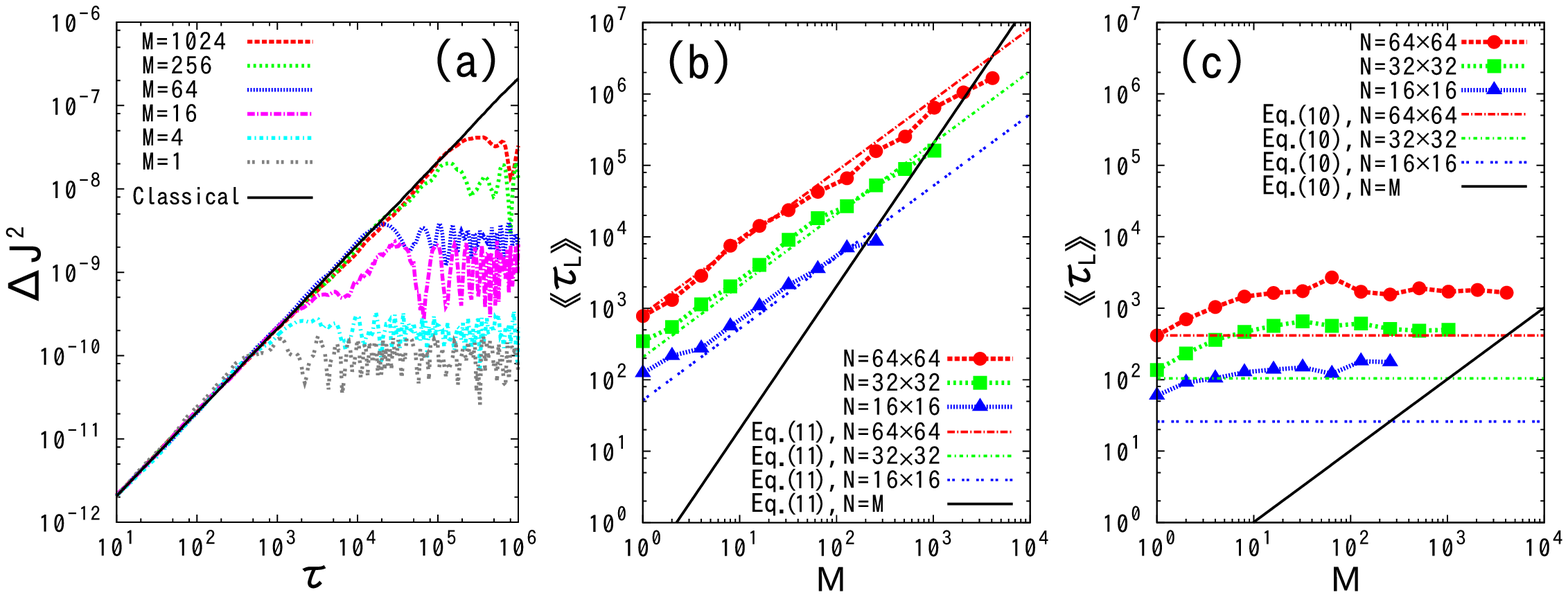}
\caption{\label{Fig3}
(a) $\msd$ vs $\tau$ for various $M$ in the regime above the threshold, where
$\eps=1.0$, $N=32\times 32$. 
(b) $\avrtlife$ vs $M$ for $\omega=\sqrt{2}$, where $N=16\times16,~32\times32,~64\times64$ denoted by 
some symbols. Theoretical results of Eq.(\ref{tlife}) are shown by lines without markers.
(c) $\avrtlife$ vs $M$ for $\omega=0$. Results of Eq.(\ref{tlife0}) are indicated by  lines without markers}
\end{center}
\end{figure*}

An advantage of our method is that it enables us to 
observe the irreversibility characteristics
even for eigenstates of $\U$ which are invariant in time except for the phase. 
Examining the characteristics of eigenstates, we encounter a remarkable
and nontrivial phenomenon. 
Let us construct the initial state
$|\Psi_0\>$ by superposing $M~(\leq N)$ eigenstates as $|\Psi_0\>=\sum_{i=1}^{M}C_m|m\>$
where 
\green{ 
$|m\>$ is the eigenstate of $\U$
} 
and $C_m\ \sim 1/\sqrt{M}$.
We show in Fig.\ref{Fig3}(a) examples of time evolution starting from such an initial state
with increasing $M$.
Apparently the \lifetime at which the $\msd$ deviates from classical diffusion increases 
with $M$, although the variation is not systematic, implying a large fluctuation of $\tlife$. 
To confirm the above 
observation we computed the average \lifetime $\avrtlife$ and show in Fig.\ref{Fig3}(b)
how it varies with $M$. The \lifetime $\avrtlife$ is proportional to $N$ if $M=1$, and 
it increases in proportion to $M$, which means $\avrtlife \propto N^2$ in the 
limit of full superposition $M=N$. Such a behavior is observed irrespective of the frequency 
$\omega$ if $\omega\neq 0$.

On the contrary, for the very particular choice of $\omega=0$, the $M$ dependence of the 
averaged \lifetime is very weak as is shown in Fig.\ref{Fig3}(c), which means that 
$\avrtlife \propto N$ for the full superposition $M=N$. 
The above results are hardly expected and they mean that
there is a basic difference between
the case of the Fourier component of quantal autocorrelation function at $\omega\neq 0$ and 
that of the Fourier component at $\omega=0$. 
We therefore consider the reason closely. 
To this end we evaluate the saturation level $\msd_\infty$ of MSD by averaging Eq.(\ref{J2corr}) 
over $\tau$.
\green{
With the use of the eigenstate $|m\>$ and its eigenangle $\gamma_m$ satisfying $\U|m\>=\e^{-i\gamma_m}|m\>$, 
a straightforward calculation yields
}
\begin{eqnarray}
\label{satu}
\nn \frac{\msd_\infty}{D}=\sum_{m=1}^M|C_m|^2 \sum_{n=1}^{N}
\frac{|\<m|\hat{v}|n\>|^2}{4}
\\ \times ~
\bigg[\frac{1}{\sin^{2}(\frac{\delta_{mn}^+}{2})}
+\frac{1}{\sin^2(\frac{\delta_{mn}^-}{2})}\bigg], 
\end{eqnarray}
where $\delta_{mn}^\pm=\gamma_m-\gamma_n\pm \omega T~({\rm mod}~2\pi)$.
Here the interference terms between 
different $m$'s are neglected (the diagonal approximation) because 
their contribution is negligibly small in the regime of concern.

First we consider the very particular case of $\omg=0$ in order to 
compare with the general case discussed later. In our setting the diagonal
element with respect to the eigenfunction 
vanishes because $|\<p|n\>|^2=|\<-p|n\>|^2$, and thus the diffusion process 
is free from the ballistic explosion beyond the Heisenberg time, which 
is well-known in the study of fidelity \cite{fidelity2}. 
Now we begin with evaluating the matrix elements. In the maximal entanglement regime $\eps\gg\eps_c$ 
all the eigenfunctions lose their identity, and so the matrix elements $|\<m|\hat{v}|n\>|^2$ between
any pair of $|m\>$ and $|n\>$ are almost the same, which allows us to approximate them by a single 
parameter, say $\<|v^2|\>$, while the expression for the autocorrelation function at $s=0$ 
is given by the relation $\corr(0)=\sum_m^M |C_m|^2 \sum_n|\<m|\hat{v}|n\>|^2 \sim N\<|v^2|\>$
(the subscript $\tau$ of $\corr$ can now be omitted), and so
\begin{eqnarray}
\label{matrixele}    \<|v^2|\> \sim \corr(0)/N . 
\end{eqnarray}
Next we consider the most important resonance factor $1/4\sin^2(\delta_{mn}^\pm/2) 
\sim 1/|\gamma_m-\gamma_n|^2$. Let us fix $m$, then in the summation over 
$n$, according to the Wigner surmise, the most dominant contribution 
comes from the nearest neighboring levels on each side, 
which have the average distance $\sim 2\pi/N$.
Therefore, the contribution from the resonance factor of each term is approximated
by $4/(4 \sin^2\pi/N) \sim N^2/\pi^2$ irrespective of $m$. With this 
and Eq.(\ref{matrixele}) we evaluate the saturation level as $\msd_\infty \sim D\corr(0)N/\pi^2$. 
The \lifetime is the time at which the classical diffusion $D_\omega^{(cl)}\tau$ is suppressed by the 
finite saturation level, namely, $D_\omega^{(cl)}\tlife=\msd_\infty$, which leads to
\begin{eqnarray}
\label{tlife0}
    \tlife \sim \frac{\corr(0)N}{\pi^2A_\omega^{(cl)}}
\end{eqnarray}
independent of $M$, where $A_\omega^{(cl)}=D_\omega^{(cl)}/D=\sum_{s=-\infty}^{\infty}\corr(s)\cos(\omega Ts)$
is the Fourier component of the classical autocorrelation function. 
This shows fairly good agreement with the numerical result 
in Fig.\ref{Fig3}(c). This time scale almost coincides with 
the so called Heisenberg time.
\\
\\
However, in the general case of $\omega\neq 0$ the \lifetime is enhanced much more 
than the case of $\omega=0$. If $\omega\neq 0$ the statistical distribution of 
$|\delta_{mn}^\pm|=|\gamma_m-\gamma_n\pm \omega T~({\rm mod}~2\pi)|$ does not suffer 
from the restriction of level repulsion in the vicinity of $|\delta_{mn}^\pm|\sim 0$. 
Hence, unlike the case of $\omega=0$, $|\delta_{mn}^\pm|$ can be arbitrarily small, 
which will make the term $1/\sin^2(\delta_{mn}^{\pm}/2)$ larger. As $M$ increases 
the chance to encounter smaller $|\delta_{mn}^\pm|$ increases.  In the summation over 
$n$ in the RHS of Eq.(\ref{satu}) the term with the smallest $|\delta_{mn}^\pm|$ and
$2\pi-|\delta_{mn}^\pm|$ will be most dominant. We evaluate the most dominant contribution.
Firstly, we fix $m$, and we take a hypothesis that 
$w_n=|\delta_{mn}^+|~(1\leq n \leq N$ and $ n\neq m)$ ( and also $w_n=|\delta_{mn}^-|$ )
are independent stochastic variables uniformly distributed over the range $[0,2\pi]$. 
The independency is a rather bold hypothesis considering that the level repulsion exists 
if $\delta_{mn}^+ \sim \delta_{mn^{'}}^+$, but its effect is quite limited.

\green{ 
Under the above hypothesis, the probability that a $w_i=w$ is the minimum in the set 
$\{ w_1,w_2,...,w_{N}, 2\pi-w_1,2\pi-w_2,...,2\pi-w_{N} \}$ is the probability
that $w_i=w$ is in $0 \le w \le \pi$ and simultaneously all 
$w_j~(j \neq i)$ satisfy $w \leq w_j \leq 2\pi-w$, which is evaluated as
\beq
\nn \frac{1}{(2\pi)^{N}} \prod_{j\neq i} \int_{w}^{2\pi-w}dw_j=\frac{(2\pi-2w)^{N-1}}{(2\pi)^{N}}
\eeq 
for $0\le w \le \pi$, which is equal to the probability of $2\pi-w_i$ being minimum and 
taking the value $w$. 
Thus the probability that the minimum value $\min\{ w_1, ..., w_{N},2\pi-w_1,...,2\pi-w_{N}\}$
takes $w$ is $p(w)=2N(2\pi-2w)^{N-1}/(2\pi)^{N}$ which asymptotically approaches 
$N\e^{-Nw/\pi}/\pi$ in the limit of $N \to \infty$. 
Secondly, we have to take the summation over $m$. At this second stage it is quite plausible to 
suppose that $w_m$ is now a statistically independent variable; then after similar evaluation as
the above it follows that the probability of $\min\{w_1,...,w_M\}$ taking a value $w$ as
\beq
  \nn  P(w) = M p(w)\bigg[\int_w^\infty p(w')dw' \bigg] ^{M-1}=\frac{MN}{\pi}\e^{-\frac{MN}{\pi}w}
\eeq
Thus the average of the minimal $|\delta_{mn}^\pm|$ is $\pi/MN$ and the most dominant 
term of $1/4\sin^2(\delta_{mn}^\pm/2)$ in the RHS of Eq.(\ref{satu}) is $(MN/\pi)^2$. Then by using
Eq.(\ref{matrixele}) and $|C_m|^2\sim 1/M$ we may evaluate  $\msd_\infty=2D\corr(0)MN/\pi^2$. 
The lifetime is decided by $D_\omega^{(cl)}\tlife=\msd_\infty$ as before to yield
\begin{eqnarray}
\label{tlife}
        \tlife  \sim  \frac{2\corr(0)MN}{\pi^2A_\omega^{(cl)}}.
\end{eqnarray}
} 
This is our final result and it agrees well with the numerical results as depicted in Fig.\ref{Fig3}(b).
In the short limit of correlation time i.e., 
$\corr(s)=\delta_{s0}\corr(0)$, Eq.(\ref{tlife})
becomes very simply $\tlife\sim 2MN/\pi^2$, independent of the details of the system. 
We have confirmed all the results discussed above are valid also for the coupled 
quantum standard maps. 
We here note that the lifetimes represented by Eqs.(\ref{tlife0}) and 
(\ref{tlife}) contain only parameters of the object system except 
for the frequency $\omega$.
We can expect that these are general features of ideally 
chaotic quantum maps which are bounded in a finite region of 
phase space.

In conclusion, we proposed a method to observe 
a characteristic of time-irreversibility, namely the decay of 
correlation over an extremely long time scale, 
in quantum systems and the associated quantum states.
Applying it to coupled kicked rotors,
the \lifetime of correspondence with classical irreversible behavior
is measured in the full entanglement regime, 
and it is found to be proportional to the number of eigenstates composing 
the examined state. In the case of full superposition of all eigenstates,
it is proportional to the square of Hilbert space dimension and is much longer than
the conventional Heisenberg time, 
which comes from the basic difference in the asymptotic behavior of
finite time Fourier components of quantum autocorrelation at non-zero 
frequency compared to the component at zero frequency.
This lifetime, in principle, should be observable 
directly in time-reversal fidelity experiments 
with an extremely weak perturbation (and so a very
high precision is required for numerical computation).

So far, the localization length has been taken as the order parameter of the 
transition to irreversible diffusion in infinitely extended quantum chaos systems 
such as the standard map. However, it can not be used in finitely bounded quantum systems. 
Our perspective is to use the \lifetime measured 
by our method for the study of transition phenomena related to the time-irreversibility 
in bounded quantum systems \cite{matsuifull} as demonstrated in Fig.\ref{Fig1}, 
combining with the finite-size scaling analysis of the critical behavior.

We finally comment that the linear oscillator L can 
be replaced by a two level system without any essential modifications, 
which means that experimental implementation could be possible 
with an optical lattice.

This work is partly supported by Japanese people's tax via JPSJ KAKENHI 15H03701,
and the authors would like to acknowledge them.
They are also very grateful to Kankikai, Dr.S.Tsuji, and  Koike memorial
house for use of the facilities during this study.


\begin{thebibliography}{0}



\bibitem{prigogine}
	\Name{Prigogine I.}
	\Book{From Being to Becoming: Time and Complexity in the Physical Sciences}
    \Publ{Freeman, San Francisco}
    \Year{1980}.



\bibitem{casati}
	\Name{Casati G. {\it et al}}
	\Book{Stochastic Behavior in Classical and Quantum Hamiltonian Systems}
	, edited by Casati G. and Ford J., Lecture Notes in Physics Vol.93
	\Publ{Springer-Verlag, Berlin}
	\Year{1979}, p. 334.



\bibitem{fishman} 
	\Name{Fishman S. {\it et al}}
	\REVIEW{Phys. Rev. Lett.} {49}{1982}{509}.



\bibitem{ikedadissip}
	\Name{Ikeda K.}
	\REVIEW{Ann. Phys.} {227}{1993}{1}.



\bibitem{cohen}
	\Name{Cohen D.}
	\REVIEW{Phys. Rev. Lett.} {82}{1999}{4951};
	\Name{Cohen D.}
	\REVIEW{Ann. Phys.} {283}{2000}{175}.



\bibitem{quantumcomputation}
	\Name{Nielsen M. A. and Chuang I. L.}
	\Book{Quantum Computation and Quantum Information}
	\Publ{Cambridge University Press}
	\Year{2001};
	\Name{C. Miquel {\it et al}}
	\REVIEW{Phys. Rev. Lett.} {78}{1997}{3971};
	\Name{G. Benenti {\it et al}}
	\REVIEW{Phys. Rev. Lett.} {87}{2001}{227901}.



\bibitem{quantumclassical}
	\Name{Adachi S. {\it et al}}
	\REVIEW{Phys. Rev. Lett.} {61}{1988}{659};
	\Name{Gadway B. {\it et al}}
	\REVIEW{Phys. Rev. Lett} {110}{2013}{190401}.



\bibitem{casatianderson}
	\Name{Casati G. {\it et al}}
	\REVIEW{Phys. Rev. Lett.} {62}{1989}{343}.



\bibitem{delande}
	\Name{Lemarie G. {\it et al}}
	\REVIEW{Europhys. Lett.} {87}{2009}{37007};
	\Name{Lopez M. {\it et al}}
	\REVIEW{Phys. Rev. Lett.} {108}{2012}{095701};



\bibitem{timerevexp}
	\Name{Ikeda K.}
	\Book{Quantum Chaos}, edited by Casati G. and Chirikov B. V. 
	\Publ{Cambridge Univ. Press}
	\Year{1996}, p.145;
	\Name{Yamada H. S. and Ikeda K. S.}
	\REVIEW{Phys. Rev. E} {82}{2010}{060102(R)}.



\bibitem{timerevexp-ballentine}
	\Name{Ballentine L. E. and Zibin J. P.}
	\REVIEW{Phys. Rev. A} {54}{1996}{3813}.



\bibitem{benenti}
	\Name{Benenti G. and Casati G.}
	\REVIEW{Phys. Rev. E} {79}{2009}{025201(R)}.



\bibitem{fidelity1}
	\Name{Peres A.}
	\REVIEW{Phys. Rev. A} {30}{1984}{1610}.



\bibitem{fidelity2}
	\Name{Gorin T. {\it et al}}
	\REVIEW{Phys. Rep.} {435}{2006}{33};
	\Name{Ph. Jacquod and C. Petitjean}
	\REVIEW{Advances in Physics}{58}{2009}{67}.



\bibitem{bermanehrenfest}
	\Name{Berman G. P. and Zaslavsky G. M.}
	\REVIEW{Physica A} {91}{1978}{450}.



\bibitem{shepe83}
	\Name{Shepelyansky D. L.}
	\REVIEW{Physica D} {8}{1983}{208}.



\bibitem{couplerotorentangle}
	\Name{Lakshminarayan A.}
	\REVIEW{Phys. Rev. E} {64}{2001}{036207};
	\Name{Fujisaki H. {\it et al}}
	\REVIEW{Phys. Rev. E}{67}{2003}{066201};
	\Name{Demkowicz-Dobrzanski R., and Kus M.}
	\REVIEW{Phys. Rev. E}{70}{2004}{066216}.


\bibitem{matsuifull}
	\Name{Matsui F. {\it et al}}
	unpublished.


\end{thebibliography}
\end{document}